\documentclass[manuscript]{geophysics}

\usepackage{amsmath}
\usepackage{graphicx}
\usepackage{amssymb}
\usepackage{mathrsfs}
\usepackage{lineno}
\usepackage{hyperref}
\usepackage{algorithm}
\usepackage{algpseudocode}
\title{ML-misfit: \\\ learn a robust misfit function for full-waveform inversion using machine learning} 
\author{Bingbing Sun and Tariq Alkhalifah}
\address{
bingbing.sun@kaust.edu.sa, tariq.alkhalifah@kaust.edu.sa \\
Physical Sciences and Engineering \\
King Abdullah University of Science and Technology \\
Thuwal, 23955, Saudi Arabia
}
\begin{document}
\maketitle 
\newpage
\begin{abstract}
	Most of the available advanced misfit functions, including those that utilize a matching filter or the optimal transport measure, are all hand-crafted. Tuning and improving those methods requires profound mathematical and physical knowledge from FWI practitioners. Besides, the performance of those misfit functions is data-dependent, and it is challenging to find a solution when these approaches fail to invert a specific dataset. We propose to learn a robust misfit function for FWI, entitled ML-misfit, based on machine learning. Inspired by the recently introduced optimal transport of the matching filter (OTMF) objective, we design a specific neural network (NN) architecture for the misfit function in a form similar to comparing the mean and variance for two distributions in OTMF. To guarantee the resulting learned misfit is a metric, we accommodate the symmetry of the misfit with respect to its input and a Hinge loss regularization term in the meta-loss function to satisfy the "triangle inequality" rule. In the framework of meta-learning, we train the network by running FWI to invert for randomly generated velocity models and update the parameters of the NN by minimizing a meta-loss, which is defined as the accumulated difference between the true and inverted models. The learning and improvement of such an ML-misfit are automatic, and the resulting ML-misfit is data-adaptive. We first illustrate the basic principles behind the ML-misfit for learning a convex misfit function for travel-time shifted signals. Further, we train the NN on 2D horizontally layered models, and we demonstrate the effectiveness and robustness of the learned ML-misfit by applying it to the well-known Marmousi model. 
\end{abstract}

\section{Introduction}
	Full-waveform inversion (FWI) is an advanced seismic imaging method. In contrast to the conventional ray-based methods \citep{tomo1}, by taking into account travel-times, amplitude, and phase information together, FWI, in principle, can deliver a model of the subsurface with resolution up to half of the propagated wavelength \citep{bleistein2013mathematics}. This feature makes FWI an essential tool for velocity model building in areas with complex geology. 
	
	FWI is generally described as a local non-linear data-fitting procedure, and it is usually solved using gradient-descent based algorithms. Given an initial model with reasonable accuracy, FWI improves the initial estimation by minimizing the mismatch between the predicted and the measured seismography. The predicted data are simulated using a wave equation solver based on an assumed physical model of the Earth. The gradient, often forming the main component for updating the model, is formulated by zero-lag cross-correlating the source generated forward-propagated wavefield and the backward propagated wavefield from receiver sides with the data residuals as excitation sources. This adjoint-state based update was initially proposed by \cite{FWI_class1} and \cite{FWI_class2}, and then subsequently many variations of FWI formulations and implementations appeared, such as frequency-domain FWI \citep{FWI_class3,FWI_class4}, Laplace-domain, and Laplace-Fourier-domain FWI \citep{FWI_class5,FWI_class6}. Together with the emergence of super-computing, improved acquisition systems (full-azimuth, ocean-bottom node, longer offsets, and lower frequency), FWI saw incredible resurgence including utilizing reflections to invert for the low-wavenumber background information, incorporating elasticity, attenuation, and anisotropy in the physical model to address more complex wave phenomena, resulting in subsurface models with improved accuracy and less uncertainty. There are many successful applications of the FWI on field data, covering the marine datasets as well as the land data ones \citep{FWI_APP_1,FWI_APP_3,FWI_APP_4,FWI_APP_5,FWI_APP_6,FWI_APP_7}. 
	
	As common to any non-linear optimization problem, FWI is prone to converge to a local-minimum rather than a global one. Especially, as FWI is trying to fit oscillating seismic signals, such a local minimum issue becomes more severe when the initial model is far from accurate, or there is limited low-frequency content in the signal. In FWI, we refer to this phenomenon as the cycle-skipping problem. A typical solution for mitigating the cycle-skipping in FWI is to build a more accurate initial velocity model or acquire ultra-low frequency data set. However, current velocity model building technology highly relies on ray-based methods and in complex regions, such as salt bodies and near-surface with topography, ray-based method can hardly recover an accurate velocity model for further FWI processing. 
	
	A cost-effective solution to reducing cycle skipping is to design a more robust misfit function to allow the model to converge to the global minimum gradually, from a far away cycle-skipped initial model. Conventional $L_2$ norm misfit functions and its variations with trace energy normalizations are widely used for its simplicity, high resolution, and robustness with respect to noise. However, its inherent local sample-by-sample comparison makes it highly sensitive to the initial model and is severely prone to the cycle-skipping problem. Thus, designing a robust misfit for FWI is the key to a successful application of FWI, and this subject has drawn a number of investigations. Recently, a series of more advanced misfit functions were proposed; these include wavefield reconstruction inversion \citep{WRI,EWI}, matching-filter based misfit \citep{Luo91,Warner,MF_Radon}, the optimal transport approach \citep{Opt_cls1,Met2016,MF_OTMF}. Specially, the optimal transport of the matching filter misfit (OTMF) was introduced by combing the matching filter and the optimal transport method together \citep{MF_OTMF}.
	
	All those mentioned misfit functions are hand-crafted. In principle, they try to compare the seismic data in a more global way to avoid the cycle-skipping issue. Such a global comparison is achieved by incorporating deconvolution and the Wasserstein distance. Though these methods have been applied successfully, their form is fixed, and they may fail for specific datasets. Improvement of their performance tends to be difficult as it requires an in-depth understanding of such algorithms, as well as the data. In this paper, we seek to learn a misfit function, entitled the ML-misfit, for FWI using machine learning (ML). 	
	Due to recent advancement in high performance computing (e.g., graphics processing units: GPUs), novel neural networks (NN) (such as the convolutional neural network (CNN) \citep{cnn}, the recurrent neural network \citep{LSTM,GRU}, etc.), as well as advanced ML optimization algorithms (e.g., the Adam method \citep{ml_adam}), the ML-based artificial intelligence (AI) algorithms have demonstrated their potential in solving complex problems, sometimes beyond the capability of humans and conventional algorithms \citep{big5_2015,big2_2016,big3_2017,big1_2019}. In the geoscience community, there is also a boom of research interest in ML and the application of ML algorithms, specifically to seismic exploration. For example, NN are trained to perform facies analysis \citep{ml_face1,ml_face2,Liu_faces_2018,ml_face3}, seismic event or first arrival picking \citep{Zhu_picking_2018,Sun_QC_2018,Shan_micro_picking_2019,hu_picking_2019}, fault detection \citep{Xiong_fault_2018,Qi_fault_2019,Wu_fault_2019}, salt body interpretation \citep{Haibin_salt_2018,Di_auto_interp_2019,Morris_auto_interp_2019,Ye_salt_pick_2019}, de-noising \citep{Dong_denoising_2019,Sun_denoising,Wu_denoising_2019,Yu_denoising_2019,Zu_deblending_2019,Li_demultiple_2020}, interpolation \citep{Jia_interpolation_2017,Jia_interp,wang_interp_2019,Wang_interpolation_2019}, 4D monitoring \citep{Musa_4D_FWI_2019,Liu_time_lapse_2019,Yuan_time_lapse_2019}, acquisition optimization \citep{Chama_acquisition_2019,Jiang_acqusition,Nakayama_survey_design_2019} and geo-engineering applications \citep{Gu_enigeer_2018,Terry_post_inversion}. 
	
	With respect to subsurface velocity model building, there are also numerous attempts to apply ML methodologies. Though the topic seems new, using NN to perform seismic inversion has a long history. \cite{nn_class_1991} trained the NN with a Monte-Carlo optimization technique to infer the depth of a seismic interface and the layer velocity from recorded arrival-times. \cite{nn_class} utilize a NN to accept a synthetic common shot gather to compute the corresponding one-dimensional velocity model. \cite{nn_langer_1996} use one station of seismographs as input to the NN, and invert for the parameters of the source, the layer thicknesses, the seismic velocities, and the attenuation along the path. \cite{nn_cal_1998} trained an NN to map the CMP gathers to interval velocities by minimizing a loss function defined in terms of the NMO-corrected data. \cite{nn_Nath_1999} perform travel time tomography using neural networks combined with a genetic algorithm and evolutionary programming techniques. \cite{nn_baronian2009} tries to invert for a dipping layered model from picked travel times using NN. These early studies were promising, but the use of supervised learning with shallow, fully connected neural networks limited their performance and their ability to scale up in size for real seismic data and applications. 
	
	Recently, \cite{Maru_fwi_tomo_2018} showed the potential of mapping the semblance of the record data to gridded velocity models with deep NN. Inspired by the information flow in semblance analysis, \cite{Gab_cmp_inversion_2020} replaced the semblance analysis with a deep CNN, and the velocity analysis is performed with a recurrent neural network (RNN). \cite{Min_velocity_analysis_2020} uses the entire semblance as well as its subfigure in a specific range as inputs to a deep CNN that allows the trained model to focus on a specific time as well as consider the overall trend of the semblance. \cite{Wu_velocity_mapping_2018} and \cite{Yang_inversion_2019} consider multi-shots seismic records as multi-channels for the input to the CNN to invert for the velocity model directly. In order to fulfill the requirement of a large dataset for training, 
\cite{deep_gan_2019} uses generative adversarial networks (GANs) to generate a geologic representation from a finite number of model examples. \cite{Wang_velocity_mapping_2019} create velocity models corresponding to a large number of natural images collected from online repositories.

The aforementioned ML-based velocity inversion algorithms try to find the inverse function to transform the data to the model directly. We refer to this category of methods as a direct velocity inversion method by ML. Another category, which we refer to as indirect methods, tries to improve or replace part of the velocity analysis or FWI steps using ML. For example, \cite{Oleg_FWI_2019} perform low frequency extrapolation in the frequency domain using common shot records as input. In their approach, the high passed and low passed record pairs are used for training a CNN to map high frequencies to low ones. After training, and still within the framework of FWI, they replace the adjoint source with ML-learned low-frequency signal, and as a result they inject crucial low wavenumber components into the model. \cite{low_frequency_sun_2018,low_frequency_sun_2019} proposed a similar CNN-based low-frequency extrapolation, but in the time domain. \cite{Jin_low_2018} and \cite{Hu_low_2019} guide the CNN training using a beat-tone generated dataset together with the raw waveform dataset. All those approaches try to use ML to help mitigating the cycle-skipping in FWI by extrapolation to the low frequencies. 

	Regularization is also an essential aspect of FWI. Using NN, \cite{reg_nn} trained a prior model with salt bodies from the migration image for regularizing the FWI. \cite{Wu_patch_cnn_2019} proposed to regularize FWI by online learning a CNN re-parametrized model for improved convergence. \cite{Gao_salt_autoencoder_2019} encoded the salt body model using an auto-encoder to perform FWI in a reduced model space and achieved better recovery of the long-wavelength salt model. 
	
	Though the NN is considered to be a universal approximator \citep{Cybenko1989}, the ill-posedness of the inverse problem makes it hard to find the mapping from the data to the model directly. Also, the increasingly large amount of seismic datasets make the direct ML approach infeasible in practical applications. Thus, the aforementioned indirect, ML-aided FWI might be more promising. Our work follows this logic and proposes to replace the hand-crafted misfit function in FWI by a machine-learned one. In our approach, we utilize the meta-learning techniques in ML to fulfill this goal. Meta-learning is a kind of semi-supervised ML technology \citep{meta_learning,meta_learning2}. It is flexible in solving learning problems and tries to improve the performance of existing learning algorithms or to learn (extract) the learning algorithms themself. It is also referred to as "learning to learn". Based on the concept of meta-learning, we can optimize the architecture of the neural network (NN) \citep{meta_search}; find the optimal initial value for the NN \citep{meta_ini}; learn an optimization algorithm \citep{meta_opt}; or learn a better loss for efficient training \citep{meta_loss}. Previously, using meta-learning, we have tried to learn an optimization algorithm, such as those mimicking the L-BFGS algorithm for fast convergence of FWI \citep{meta_gradient}. 

	In this paper, we further expand the application of the meta-learning methodology in FWI and try to learn a robust misfit function (distance, metric). In computation science (CS), metric (distance) learning is a branch of machine learning that aims to learn distances from data. The metric learning method developed in CS can be used to improve the classification and data reduction \citep{metric_learning}. Thus, the methodology proposed differs from the classic metric learning problem by learning a metric (distance, misfit) to improve the convergence of an optimization problem, and especially for FWI. Similar attempts along this line can be found in the Arxiv preprint website\citep{meta_opt}, and they use meta-learning to learn a loss for better training of the classification and the reinforcement learning problem. In their approach, the learned loss is a black-box represented by an NN and might not be a metric defined in the mathematics (needs to comply with several rules, such as the triangle inequality). Moreover, inspired by the hand-crafted misfit, the NN architecture for our misfit function is designed to be of a specific form. A specifically designed architecture makes the resulting misfit function admit some critical properties for a metric, as well as the global comparison features for a cycle-skipping free objective. 
	
	In summary, the ML-misfit proposed in this paper is specific to FWI and has the following features:
	\begin{itemize}
	\item Uniform framework:    We formulate the general meta-learning framework for learning a misfit in FWI;
	\item Unique NN architecture: Rather than formulate the misfit function as a fully black-box NN, we design a specific NN architecture for the ML-misfit, which in principle mimics reducing the mean and variance of the resulting matching filter distribution in the OTMF approach \citep{MF_OTMF}; 
	\item Metric-learning: To ensure the resulting ML-misfit is a metric (distance) in mathematics, we introduce symmetry to the NN of the misfit function and a Hinge-loss regularization term in the training steps. 
	\end{itemize}
	
	In the following, we have a brief review of the FWI technology and then describe the meta-learning framework for learning the ML-misfit for FWI. We will detail the NN architecture and the training procedure of the meta-learning. In the examples section, we demonstrate the success of learning a convex misfit for time-shifted signals. Then we present the results of training the NN on a 2D layered model and its inversion result of the Marmousi model. At last, we include discussions.
		
\section{Theory of Full Waveform Inversion}
	FWI tries to find a parameter model $\mathbf{m}$ that can be used to produce predicted (synthetic) data $\mathbf{p}$ that is close to the measured data $\mathbf{d}$. Assuming that the waves propagating in the subsurface obey certain physical laws (such as acoustic, elastic, viscoelastic, etc.), the predicted data $\mathbf{p}$ is simulated use a proper forward modeling algorithm, that is extracted from the wave equation operator $\mathcal{L}$ to generate the predicted data $\mathbf{p}(\mathbf{m})$:
 \begin{equation}
 \mathcal{L}(\mathbf{m})= \mathbf{p}.
 \label{eqn_forward}
 \end{equation}
 FWI is formulated as a nonlinear optimization (inverse) problem to invert for the model parameter $\mathbf{m}$. Thus, we define an objective (misfit, loss) function $\Phi$, which takes the measured data $\mathbf{d}$ and the predicted data $\mathbf{p}$ as input and outputs a scalar value. The purpose of such misfit is to measure the dissimilarities between $\mathbf{p}$ and $\mathbf{d}$. The property of the objective function $\Phi$ will effect the inversion, and normally we prefer a convex misfit function (with respect to the time shift) for its ability to reduce cycle-skipping. 
 The solution to this inverse problem can be expressed as :
 \begin{equation}
 \mathbf{m}' =\underset{\mathbf{m}}{\operatorname*{arg\,min}} {\Phi}\left[\mathbf{p}(\mathbf{m}),\mathbf{d}\right],
 \end{equation}
 where $\mathbf{m}'$ is the optimal model parameter leading to a minimum value of the misfit function $\Phi$.
 We compute the derivative of the misfit function $\Phi$ with respect to the predicted data $\mathbf{p}$, which is referred to as the adjoint source:
 \begin{equation}
 \mathbf{s} =  \frac{\partial \Phi}{\partial \mathbf{p}},
 \end{equation}
 By backpragating the adjoint source, which is equivalent to a reverse time migration (RTM) with the adjoint source as the record, we obtain the gradient of the misfit function with respect to the model:
 \begin{equation}
 \mathbf{g}_k = \text{RTM}(\mathbf{s}_k).
 \end{equation} 
 To update the model along with descent-direction, we have:
 \begin{equation}
 \mathbf{m}_{k+1} = \mathbf{m}_{k} - \gamma_k \mathbf{g}_k,
 \label{RTM}
 \end{equation}
 where $k$ denotes the FWI iteration step, $\gamma_k$ is the line-search step. We perform those steps iteratively till we reach a threshold value for the misfit value (or the gradient is smaller than a predefined value) or a predefined maximum iteration number is met. 
 
 The selection of the misfit function $\Phi$ is important for a successful application of FWI, and there are many options for choosing such misfits, among which the simplest one is the (squared) $l_2$ norm misfit. Suppose the predicted data $\mathbf{p} = p(t)$ and $\mathbf{d} = d(t)$, the $l_2$ norm misfit function can be defined using a sample-by-sample subtraction between the predicted and the measured data:
 \begin{equation}
 \Phi_{\text{L2}}\left(p,d\right)= \frac{1}{2} ||p(t)-d(t)||_2^2,
 \label{misfit_l2}
 \end{equation}
 where the $l_2$ norm misfit here is defined for a single trace of predicted and measured data as input, and we omit the summation over sources and receivers for brevity. Our ML-misfit will adhere to such definition, i.e., working on a single trace rather than a common record. This makes the ML-misfit methodology more adaptable to large scale problems. In the next section, we will describe how to design the NN architecture of the ML-misfit and the procedure of training the NN based on the framework of meta-learning. 
 
\section{meta-learning for a misfit function in FWI}
\subsection{The architecture for the neural network of the ML-misfit}
In ML-misfit, the misfit function will be parameterized by an NN. We can consider the NN as a black box and do not assign a specific form to the network. For example, we can define the NN as having the predicted and measured data as input and a scalar value as output to represent the misfit. Though such design is simple and straightforward, a black-box approach is not optimal as the NN has to rely on the training step to learn everything needed to become a proper misfit, which makes the training complicated. 
Thus, to better constrain the function space and stabilize the training of the NN, we suggest the following NN architecture for the ML-misfit $\Phi_{\text{ML}}(p,d;\theta)$:
\begin{equation}
\Phi_{\text{ML}}(p,d;\theta)  = \frac{1}{2}|| {\phi}(p,d;\theta)- {\phi}(d,d;\theta)||_2^2 + \frac{1}{2}||\phi(d,p;\theta)- \phi(p,p;\theta)||_2^2,
\label{ML_misfit_form}
\end{equation}
where $\phi(p,d;\theta)$ represents an NN with the inputs $p$ and $d$ in vector form (single trace), and its output is a vector, as well, but of 2 values. Here, $\theta$ represents the NN parameters, which we will train later. The form of the ML-misfit in Equation \ref{ML_misfit_form} is inspired by the OTMF misfit function \citep{MF_OTMF}. In the appendix, we include a brief review of the OTMF methodology and discuss the inspiration to form equation \ref{ML_misfit_form}. 

The misfit function of Equation \ref{ML_misfit_form} consists of two terms. Let us focus on the first term: $\Phi_1= || {\phi}(p,d;\theta)- {\phi}(d,d;\theta)||_2^2$ (see Figure \ref{fig:phi}) as the second term $\Phi_2=||\phi(d,p;\theta)- \phi(p,p;\theta)||_2^2$ is introduced to guarantee symmetry of the misfit function with respect to the inputs (with the second term, $d$ and $p$ are just switchable for the ML-misfit, i.e., $\Phi_{\text{ML}}(p,d;\theta)=\Phi_{\text{ML}}(d,p;\theta)$). The network $\phi$ with parameters $\theta$ takes two traces of data as input $(p,d)$ or $(d,d)$, and their outputs are given a two-dimensional vectors, which are expected through intuition to be similar to the mean and variance, like in the OTMF approach. We assume that evaluation of ${\phi}(d,d;\theta)$ admits zero mean and variance values considering the two input traces $(d,d)$ are identical. In contrast, ${\phi}(p,d;\theta)$ outputs different mean and variance values indicating the dissimilarity between the two input traces of $p$ and $d$. Of course, such outputs of the network reflect our intuition that the NN might simulate the OTMF objective function. However, as ML-misfit adapts to the data, and the dimension of the output can be enlarged and it may represents more information beyond the mean and the variance. The (squared) $l_2$-norm of the difference between ${\phi}(p,d;\theta)$ and ${\phi}(d,d;\theta)$ will measure the departure of the predicted data $p$ from the measured data $d$, and this operation mimics the Wasserstein distance computation step in the OTMF approach. Refer to Appendix A for a detailed analysis of the connections between the ML-misfit of Equation \ref{ML_misfit_form} and the OTMF misfit. 

Courtesy of the form we introduced in Equation \ref{ML_misfit_form}, we can verify that the ML-misfit satisfies the following rules for a metric (distance):
\begin{eqnarray}
\Phi_{\text{ML}}(p,d) \ge 0, \\\
\Phi_{\text{ML}}(f,f) = 0  \Leftrightarrow f=f, \\\       
\Phi_{\text{ML}}(p,q) = \Phi_{\text{ML}}(q,p),    
\label{metric_1}
\end{eqnarray}
where $p,d,f,q$ are arbitrary input vectors and from this point on, we omit the ML-misfit dependency on the NN parameter $\theta$ for conciseness. The remaining important requirement for a metric is the "triangle inequality" rule:
\begin{equation}
\Phi_{\text{ML}}(p,q) \le \Phi_{\text{ML}}(p,n)  + \Phi_{\text{ML}}(n,d), 
\label{metric_2}
\end{equation}
where $n$ is an arbitrary input vector. The ML-misfit given by Equation \ref{ML_misfit_form} does not fulfill the "triangle inequality" rule automatically. Thus, we introduce a Hinge loss regularization to comply with the condition:
\begin{equation}
R_{\text{HL}}( p, q, n )= \max( 0,\Phi_{\text{ML}}(p, q) -\Phi_{\text{ML}}(p,n) - \Phi_{\text{ML}}(n,q) ).
\label{hinge}
\end{equation}
If the "triangle inequality" rule of Equation \ref{metric_2} holds for the ML-misfit, the Hinge loss function of Equation \ref{hinge} would be zero. Otherwise we seek a minimization of Equation \ref{hinge} to force the resulting ML-misfit to comply with the "triangle inequality" rule. We will detail the application of such Hinge-loss regularization in the next section dedicated to training of the ML-misfit. In the example section, we will demonstrate that the NN architecture of Equation \ref{ML_misfit_form}, as well as the Hinge loss regularization, improve the training speed and the convexity of the resulting learned ML-misfit. 

We can also notice that Equation \ref{ML_misfit_form} can reduce to the $l_2$ norm misfit, by setting $\phi(x,y;\theta)=x$. This potentially suggests that the machine, if it can, might learn a cycle-skipping free objective and, at the same time, adapts to the high-resolution features of the $l_2$ norm misfit. 

We use CNN for the neural network $\phi$, taking the predicted data $p$ and $d$ per trace as input. After each CNN, we use LeakyRelu activation function and then a MaxPoolling layer to reduce the size of the data to half. After several convolution layers, we add a fully connected layer whose output mimics the statistics of the distributions in the OTMF approach, and we set the size of the output to be 2 (mimicking the mean and variance). We will describe the details for parameters of NN for specific problems in the examples section. 

\subsection{ Training the neural network}
In this section, we describe how to train the NN of the ML-misfit defined in the previous section using meta-learning. 

In meta-learning, the training dataset is a series of tasks, rather than labeled data in supervised learning problems such as classification. Our loss for the training referred to as the meta-loss function is defined to measure the performance of the current neural network for the implementation of those tasks. Back to our ML-misfit learning problem, the tasks are formulated by running FWI applications, and the process is the same as conventional iterative waveform inversion except for replacing the $l_2$ norm misfit (or any other misfit) with a neural network, i.e., the ML-misfit. 
We run many FWI applications, e.g., using different models. In the training phase, as the true models are available, we define the meta-loss as the normalized $l_2$ norm of the difference between the true model and the inverted one plus the Hinge loss regularization, e.g., 
\begin{equation}
\text{Loss} = \frac{1}{2} \sum_{k'=k}^{k+K} \frac{||m_{k'}-m_{\text{true}}||_2^2}{||m_{\text{true}}||_2^2}, + \lambda \sum_{k'=k}^{k+K} R_{\text{HL}}( p_{k'}, d, n_{k'} ),
\label{eqn_meta_loss}
\end{equation}
where $R_{\text{HL}}$ is the Hinge loss function defined in Equation \ref{hinge}, and $\lambda$ is the weighting parameter for the Hinge loss. $K$ is the unroll integer meaning every $K$ FWI iteration we update the NN parameters. Thus, $m_{k}$ and $m_{\text{true}}$ are the inverted model at iteration $k$ and the true model, respectively. Likewise, $p_{k}$ and $d$ are the predicted data at iteration $k$ and the measured data, respectively. The data $n_{k}$ are randomly generated in the training step. To produce realistic waveforms for $n_k$, in each iteration, we generate a random number $\epsilon$ from a uniform distribution $[0,1]$, and the data $n_{k}$ is simulated from a model $m_k^n$ created by a random interpolation between the current inverted model and the true one: $m_{k}^n = (1-\epsilon)*m_{\text{true}} +\epsilon*m_{k}$.
Besides, a linear interpolation makes the resulting Hinge loss regularization term smaller when the current inverted model $m_k$ becomes closer to the true one $m_{\text{true}}$, and this is in consistent with the first term, which becomes smaller as well in this case. Thus, this strategy of generating random data $n_k$ for applying the Hinge loss function makes the selection of the weighting parameter $\lambda$ easier, and thus, stabilizes the training process. 

Note, other forms of meta-loss can also be used. For example, we can replace the model difference in equation \ref{eqn_meta_loss} with data difference or any other hand-crafted misfit function (or even, another already learned ML-misfit function). Note, the meta-loss is defined to measure the performance of the current learned ML-misfit function, as long as such meta-loss defined can verify the goodness of the inversion result, it is sufficient to perform the meta-learning. In our case, to reduce cycle-skipping, an ideal learned ml-misfit would reduce the model misfit, as well as the data misfit.

In the training, to update the NN parameters, we need to figure out the dependency of the meta-loss on the NN parameters $\theta$. Given model $m_k$ at the current iteration k, we perform the forward modeling to obtain the predicted data $p_k$. Here, for clarity, we only consider the first order dependency, i.e., we assume $m_k$ does not depend on the NN parameters of the ML-misfit and So does the predicted data $p_k$. 

The derivative of the ML-misfit with respect to the predicted data leads to the adjoint source (data residual):
\begin{equation}
s_k = \frac{\partial \Phi_{\text{ML}} (p_k,d)}{\partial p},
\label{adjoint_source_ml}
\end{equation}
Note that the resulting adjoint source is dependent on the parameter $\theta$ of the ML-misfit. 

Similar to equation \ref{RTM} but with an adjoint source computed by ML-misfit rather than the $l_2$ norm, we backpropagate the adjoint source to get the model perturbation:
\begin{equation}
g_k = \text{RTM}(s_k),
\end{equation}
and the model can be updated accordingly:
\begin{equation}
m_{k+1} = m_k - \gamma_k g_k.
\label{eqn_model_updating}
\end{equation}
Using the updated model $m_{k+1}$, we can simulate the predicted data $p_{k+1}$ for next iteration and iteratively repeat this process to update the model. In this process, the predicted data $p_{k+1}$ depends on NN parameters $\theta$ through the model $m_{k+1}$ and the model $m_{k+1}$ depends on $\theta$ through the gradient $g_k$, which further depends on the adjoint source $s_k$ through the RTM process. 

Thus, when we obtained the updated model $m_{k+1}$, the predicted data $p_{k+1}$ as well as the random signal $n_k$ generated according to our previous described procedure, we can compute the associated meta-loss of equation \ref{eqn_meta_loss} and then compute the gradient with respect to the NN parameters $\theta$ to update:
\begin{equation}
\theta_{new} = \theta -\eta \frac{\partial {\text{Loss}}}{\partial \theta},
\label{eqn_descent}
\end{equation}
where $\eta$ is the learning rate. The gradient computation is based on automatic differentiation, which is provided in most of the modern ML framework, such as "Pytorch" (which is used in our research and the implementation of the ML-misfit). Equation \ref{eqn_descent} represents a standard gradient descent updating scheme. For efficiency and accuracy, we normally adopt more advanced optimization algorithms such as Adam or RMSprop. Those algorithms incorporate moments information for an adaptive learning rate and provides faster convergence for training the neural network, especially when the network is large and deep. 

We can only measure the meta-loss at last FWI iteration step and  back-propagate the residual to update NN parameter. However, this would makes the residuals hard to propagate to earlier steps as it will becomes smaller and smaller. We will not also update the NN parameters every FWI step, as this approach is similar to using a batch size of one in supervised learning, which is usually less stable. Thus, as shown in equation \ref{eqn_meta_loss}, the meta-loss is accumulated with multiple K steps for stability consideration. This means every K steps, we will update the NN parameter by back propagating pre-K steps residuals, simultaneously. Before, we assume that the current model $m_k$ does not depend on the NN parameter. Actually, there is a high order dependency between the inverted model and the NN parameters $\theta$ (granted the iteration step is larger than 1, which is normal). Because $m_{k+1}$ depends on the NN parameters not only through the gradient $\text{RTM}(s_k)$, but also through $m_k$ (we ignore this dependency in our first-order analysis), which further depends on the model $m_{k-1}$ and $\text{RTM}(s_{k-1})$ and so on. Though these dependencies seem complex, the ML framework will handle this automatically and update the NN parameter with auto-differentiation schemes. 

Another aspect we need to point out is that the updating of the parameters $\theta$ of the NN requires dealing with high-order derivatives, i.e., the gradient of a gradient. This is because $s_k$ is already the derivative of the ML-misfit (with respect to the predicted data). Updating of the neural network requires the computation of the derivative of the adjoint source $s_k$ with respect to the parameters $\theta$. Most modern ML frameworks include modules for such high-order derivatives, e.g., in Pytorch, we can use the function "torch.autograd.grad" for dealing with this issue. 

To better understand the dependency between the models and the NN parameters, we draw a typical FWI flow using the ML-misfit in Figure \ref{fig:fwi_flow}. The gradient flow for updating the NN parameter are the blue and red curves. The red curves correspond to high order dependency as described before. We also present the pseudo code for training the ML-misfit in Algorithm 1.

In summary, the first term of Equation \ref{eqn_meta_loss} asks the ML-misfit to converge faster with the least model residuals, and this is equivalent to optimize for a cycle-skipping free objective as a cycle-skipped model will always correspond to large model residuals. The second term seeks an ML- misfit, which complies with the "triangle inequality" rule.  Considering that ML-misfit already satisfies equation \ref{metric_1}, the optimization based on equation \ref{eqn_meta_loss} will lead to a cycle-skipping free misfit function and at the same a metric (distance) in mathematics. In the example section, we will demonstrate these desirable features of the ML-misfit learned by the machine. 

\section{Examples}
We will start with simple travel-time shifted signals, beyond the cycle constraint, to teach the NN to overcome this limitation of classic misfits. Then, we use a 2D model to teach the NN to deal with more realistic nonlinearity.

\subsection{Learn a convex misfit function for travel-time shifted signals}
We normally use shifted signals to evaluate the convexity of a misfit function in FWI. Similarly, in this section, we designed a light-weight "FWI" for inverting a single travel-time, which controls the shift in the signal. Based on the proposed meta-learning framework, described in the previous section, we will teach the ML-misfit to invert such a travel-time shift efficiently. We first share an in-detail description of the setup of the experiment.

\subsubsection{Experiment setup}
In conventional FWI, given the model parameter $\mathbf{m}$, the predicted data $\mathbf{p}(\mathbf{m)}$ is simulated using a partial differential equation (PDE), and the updating of the model parameters corresponds to solving the PDE another time. This costly process makes the testing and evaluation of the ML-misfit time- consuming. Thus, in this example, we propose a simplified FWI to evaluate our method, efficiently. In this mini FWI, we only invert for a single parameter, i.e., the travel-time shift $\tau$. An assumed forward modeling produces a shifted Ricker wavelet representing the predicated data $p(t;\tau)$ :
\begin{equation}
	p(t;\tau) = \left[1-2\pi^2f^2(t-\tau)^2\right]e^{-\pi^2f^2(t-\tau)^2},
	\label{travel_time_modeling}
\end{equation}
where $f$ is the dominant frequency. Equation \ref{travel_time_modeling} leads to a fast simulation of the predicted data $p(t;\tau)$ from the travel-time shift parameter $\tau$, which significantly accelerates the testing and evaluation of our methodology. 

The meta-loss function of Equation \ref{eqn_meta_loss} for training is modified accordingly:
\begin{equation}
	\text{Loss} = \frac{1}{2} \frac{(\tau_{\text{true}}-\tau)^2}{\tau_{\text{true}}^2} + \lambda R_{\text{HL}}\left[ q(t;\tau), q(t;\tau_{\text{true}}),q(t;\tau_n)\right],
	\label{meta_loss_travel_time_shift}
\end{equation}
where we omit the summation over multiple steps for brevity (in this example, the unroll parameter $K$ is 10). As before, the travel-time shift $\tau_n$ used in the Hinge loss regularization term is computed by a linear interpolation between the true travel-time $\tau_{\text{true}}$ and the current inverted travel-time shift $\tau$: $\tau_n = (1-\epsilon)*\tau_{\text{true}} + \epsilon *\tau$, where $\epsilon$ is a random number sampling from a uniform distribution in $[0, 1]$. This makes the modeled data $p(t;\tau_n)$ a shifted signal between $p(t;\tau)$ and $p(t;\tau_{\text{true}})$. As a result, the Hinge loss regularization can take into account such time-shift features to apply the "triangle inequality" constraint. 

	In this example, we discretize the waveform of the signal using 100 samples with a sampling interval 0.02s, and thus simulate a record of length 2 s. In general, the NN architecture of $\phi$ in Equation \ref{ML_misfit_form} mimics the AlexNet neural networks \citep{AlexNet} in that it uses larger kernels at the earlier convolution layers and smaller kernel sizes with more channels at later layers. In total, we have five convolution layers and one fully connected layer. The inputs to the NN of $\phi$ are two vectors, i.e., the predicted and measured data of one trace, and they are considered as two channels for the first convolution layer. We follow each convolution layer with a LeakyRelu activation function, followed by a MaxPooling. We set the channel number to $(64,128,256,256,64)$ for the five convolution layers, while the corresponding kernel sizes for the convolution layers are set to be $(17,9,9,5,5)$ with a stride of size one. (We set a relatively large kernel size of 17 for the first convolution layer. As in the training, we use a minimum frequency of 3 Hz whose period is approximately 17 samples for our setup.) The kernel and stride size are set both to be two for the MaxPooling layer, and thus, we halve the length of the data after the MaxPooling operation. The fully connected layer will take the flattened output from the previous MaxPooling layer and output a vector with a size of two (which might mimic the mean and variance in the OTMF approach). We do not use a DropOut or a BatchNormalization layer in our application. 

%In a black-box approach, the NN takes in two vectors as input and outputs a scalar value representing the misfit. Thus, to formulate the NN of the black-box misfit within the suggested framework, we add an extra fully connected layer that takes the input of size two (which is the size of the output from the NN of $\phi$) and outputs one single value representing the misfit. In this setting, the difference in the number of the parameters between the black-box approach and the proposed NN architecture by equation \ref{ML_misfit_form} is negligible, the only addition for the black-box approach is the last fully connected layer which adds only two weights and one bias. Specially, considering that the number of the parameters for the NN $\Phi$ is large (for $nc=1$, the number of parameter is 10690), the extra addition of three parameters will not effect the fair comparison of the two NNs.

In each epoch, we randomly generate 600 true and initial travel-times between 0.4 s and 1.6 s. To inject more variety in the training, we also randomly set the main frequency in equation \ref{travel_time_modeling} between  3 Hz and 10 Hz. We invert for 60 travel-times, simultaneously. We run 100 iterations for each inversion and set the learning rate for updating the travel-time shift to one. As we update NN parameters every ten iterations (the unroll parameter $K=10$), the batch size is equivalent to be 600. We adopt the Adam algorithm for training the neural network. The learning rate is set to be a constant of 1.0e-5 and we train for 500 epochs in total and after 300 epochs, we scale the learning rate by 0.3. We create another 60 inversion problems for testing (In the testing dataset, the true travel-times, as well as the initial travel-times for starting the inversion, are also randomly generated but kept fixed during the training). 

\subsubsection{Results and discussions}
	After 500 epochs of training, we show in Figure \ref{fig:misfit}, the curve of the normalized meta-loss of equation \ref{meta_loss_travel_time_shift} over epochs for the training and testing tasks. The continuous reduction in the loss for the training and testing tasks shows convergence and demonstrates the success of the training of the ML-misfit neural network. 
	
	To evaluate the convexity of the ML-misfit with respect to the travel time shift, we compute the ML-misfit between a target signal and its shifted version with a varying time-shifts. In the computation, we choose the main frequency to be 6 Hz and the travel-time for the target signal is set to at 1.0 s, while the time-shifts with respect to the target signal varies from -0.6 s to 0.6 s. Figure \ref{fig:convexity_without_train,convexity_train_500} shows the resulting normalized misfit value for the $l_2$ norm and the trained ML-misfit over epochs. It is obvious that the non-convexity of the $l_2$ norm with respect to time shifts resulted in local minima. The ML-misfit without training has its parameters randomly initialized. Its misfit value with respect to the time shift, as shown in Figure \ref{fig:convexity_without_train}, and as expected, shows random behavior as well. However, as shown in Figure \ref{fig:convexity_train_500}, the ML-misfit after training improved its convexity with respect to travel-time shifts and finally ML-misfit results in a reasonably convex objective function without observable local minima. This indicates that the ML-misfit does learned a more robust way to compare the predicted data and measured data. 
	
	To further evaluate the accuracy of the resulting ML-misfit function, in Figure \ref{fig:test_check}, we plot the true and inverted travel-times for the testing datasets. Note in the figure, the x-axis (labeled 'index") denotes different value, and the y-axis corresponds to the travel-time. We can see that the inverted travel-time shows good accuracy even though the initial travel-time is far from the true value indicating a strong cycle-skipping issue (note that the frequency range is between 3Hz and 10 Hz). We can observe some spikes indicating less accuracy of the result. Those spike can be mitigated for improved accuracy by tuning the training parameters, adding extra regularization terms (such as sparse regularization of NN parameters) or adjusting the neural network architecture.
	This simple, while illustrative travel-time shifted signals example demonstrates that based on the proposed meta-learning framework, NN can learn a cycle-skipping free misfit. An introduction of a specific architecture for the misfit function and including a Hinge loss regularization function can improve the efficiency of the training as well as the convexity of the resulting misfit function.

	In the next section, we will increase the number of channels of the convolution layer in our NN, which helps the NN work with more complex and realistic waveforms (rather than a single event shown in this section). Specially, we will train the NN on a 2D layered model by running FWI (not a simplified version as in this section) and apply the resulting learned ML-misfit to the Marmousi model.
			
\subsection{The Marmousi model}
	We train the ML-misfit on a randomly generated 2D horizontally layered model and apply the resulting learned misfit to Marmousi model.  
	
	For training, the model size is set to be 2 km in depth and 8 km wide, with a sampling interval 40 m in both directions and we mimic a marine geologic setup with velocity ranging between 1500 and 4200 m/s. The 2D layered model is generated randomly with a general increase of velocity with depth. Specially, we randomly create a series of layers $z_{i}$ with $z_0 = 0$ km. Thus, the top of the layer $i$ is at depth $z_{i-1}$, and the layer bottom is at depth $z_{i}$. The interval velocity for layer $i$ is set to be $v_i= 1500 + 1.35\epsilon z_{i}$, where $\epsilon$ is a random number in [0, 1]. To mimic a marine step up and also to stabilize the inversion, we set the first layer of the model to have a water velocity of 1500 m/s and fix it in the inversion. The water depth for the water layer is randomly generated from 100 m to 500 m. The velocity model for FWI will be presented on a regular gird (50 by 200 here). With these semi-random velocity models, we utilize finite-difference modeling to generate the data. The initial model used to start the FWI is obtained by applying strong smoothing to the true model with a Gaussian smoother of a standard deviation of 2 km. 
	
	As the model is a horizontally layered model, we only need to compute one-shot record to perform the inversion. Thus, we set the shot on the surface with shot location at 0. We have 50 receivers evenly spaced at the surface. The recording time is 4.8 s, with a time sampling interval of 0.03 s, and thus number of samples in each trace is $4.8/0.03=160$. In training, we use a 6.5 Hz central frequency Ricker wavelet with energy below 3 Hz muted, so that the learned ML-misfit can mitigate the cycle skipping without such low frequencies. We perform the full waveform inversion on full-band data without frequency-continuation. Besides, to stablize the inversion, we add TV regularization of the model \citep{TV1,TV2} for FWI. 
	We randomly generated 64 models and kept them for testing, and in each epoch, we randomly generated 256 models for training.  
	
	The NN architecture for the neural network $\phi$ is the same as that in the time-shifted signal examples. The input size of the trace for the neural network is 160, which is the entire length of the record. To accommodate the complexity in seismic data, such as those corresponding to the Marmousi model, we increase the number of channels of the five convolution layers to $(128,256,512,512,128)$. The kernel size of each layer is modified to $(15,15,5,5,5)$ correspondingly.  
	
	As shown in Equation \ref{eqn_meta_loss}, we adopt the $l_2$ norm of the model difference and the Hinge loss regularization as the meta-loss. we update the parameters of the ML-misfit every 10 FWI iterations (unroll integer $K=10$). And thus, considering we have 50 receivers, the batch size for training is 500. 
	
	We train the NN for 100 epochs. Figure \ref{fig:marm_epoch} shows the normalized meta-loss for the training and testing tasks over iterations, the reduction of the loss value suggests the convergence of the training. Figures \ref{fig:NEW_check0False} and \ref{fig:NEW_check110False} show profiles of the inverted models selected from the testing data using the ML-misfit before and after training. We can see that the inversion of the layered models get improved after training, especially for the shallow part of the model. 
	
	We apply this trained ML-misfit to the modified Marmousi model, shown in Figure \ref{fig:marm_vtrue2}. Similarly, this Marmousi model extends 2 km in depth and 8 km in distance. We simulate 80 shots, and 400 receivers are spread evenly on the surface. As in the training, wavelet with a 6.5 Hz central frequency is used in the inversion. To demonstrate that the learned ML-misfit could mitigate the cycle-skipping without low frequency, we mute frequencies below 3 Hz. In order to use the ML-misfit trained on the horizontally layered model, the size of the input must be consistent. Thus, we simulate the record time up to 4.8 s the same as that used in the training step.
	Figure \ref{fig:marm_vini2} shows the initial model for starting the inversion. As before, we perform a full-band inversion (band-pass the data in the frequency range of 3 Hz to 7 Hz) without frequency continuation.
	Figures \ref{fig:marm_L2_misfit} and \ref{fig:marm_ML_misfit} show the inversion results using the $l_2$ norm and ML-misfit, respectively. The result using $l_2$ norm shows obvious cycle-skipping features, while the ML-misfit shows considerably improved results with an ability to recover the low-wavenumber components of the model, such as in the left part of the model.    
	\section{Discussion}
The proposed ML-misfit framework focuses on using a neural network to perform a measure of misfit (distance) between the observed and predicted data that is free of cycle skipping and adaptable to the data for waveform inversion. Using meta-learning principles, the learned ML-misfit, depending on the learning data, can perform at least like an $l_2$ norm, but potential includes features that can help us avoid cycle skipping as the meta-loss in the training inspired such a misfit, but even more so it can include additional features to mitigate other data complexities. More importantly, since the ML-misfit is performing a specific task within the waveform inversion machinery, it is, unlike many other networks based on supervised learning to directly build a velocity model, easily usable in real data. The application of the ML-misfit on the Marmousi data, which was trained on a horizontally layered model demonstrated some of this flexibility. As this paper is focused on introducing and proving the concept, we will show applications to field data in future work, as such applications require their own analysis.
	\section{Conclusion}
	We developed a framework for learning a robust misfit function, entitled ML-misfit, for FWI using meta-learning. A specific neural network architecture as well as a Hinge loss function, are used to shape the resulting ML-misfit to be a metric. We demonstrate the basic principle and the ability of ML-misfit to learn a convex function, such as for simple travel-time shifted signals. Trained on a randomly generated 2D layered models, the resulting ML-misfit can invert for the Marmousi model free of cycle-skipping issues using a signal without frequencies below 3 Hz.  
%\section{Generalization and extension for the ML-misfit}
%	Though we present our work here in a specific form, our proposed invention provides a general framework for learning a misfit function in inverse problems. The %options for generalization are many and here we provide examples:
%	\begin{itemize}
%	\item We defined the NN architecture to be of specific form in equation 1. A more general form of the NN architecture can be set as a black box function:
%	\begin{equation}
%	J_{\text{ML}}(p,d) = \Phi(p,d;\theta),
%	\end{equation}
%	we will not introduce any symmetry and let the machine learn on its own. 
%	\item We use DCN for the NN $\Phi$ in the above example, but conventional neural networks (CNN) or recurrent neural network (RNN) can also be options. 
%	\item We use a shallow network for the NN $\Phi$, deeper networks using the ResNet framework can be utilized for improving the accuracy and robustness of the resulting ML-misfit. 
%	\item The input to the ML-misfit network described here is a 1D trace signal, other ensembles of data can also be used such as common shot, common receiver, common mid-point, common azimuth or any other combinations.
%	\item The input to the ML-misfit network described here is in time domain, other transformed domains of signal can also be used, such as time-frequency domain, Fourier domain, Wavelet domain, Radon domain etc.
%	\item The training of the NN of the ML-misfit discussed above is based on meta-learning, we can also use reinforcement learning for training such NN. 
%	\end{itemize} 
\onecolumn
\bibliographystyle{seg}  % style file is seg.bst
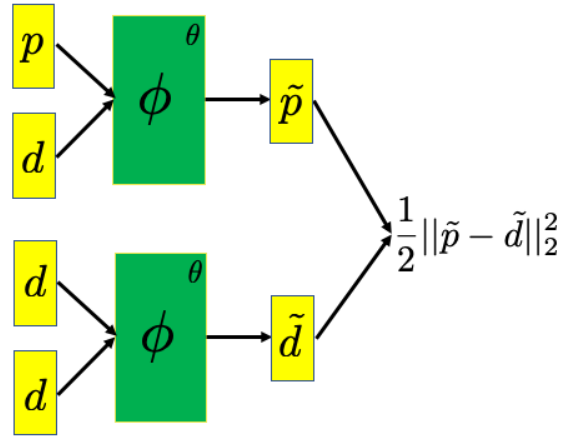
\plot{phi}{width=0.49\textwidth}{A schematic plot of the neural network architecture of the ML-misfit. The predicted data $p$ and measured data $d$ correspond to a single trace of the seismic data in vector form. They are packed to form two inputs, i.e., $(p,d)$ and $(d,d)$. Those inputs go through the neural network $\phi$ with parameters $\theta$, and outputs $\tilde{p}$ and $\tilde{d}$, respectively. The resulting misfit function is defined to be the $l_2$ norm of the difference between the outputs of $\tilde{p}$ and $\tilde{d}$. See the section regarding the neural network architecture for more information. }
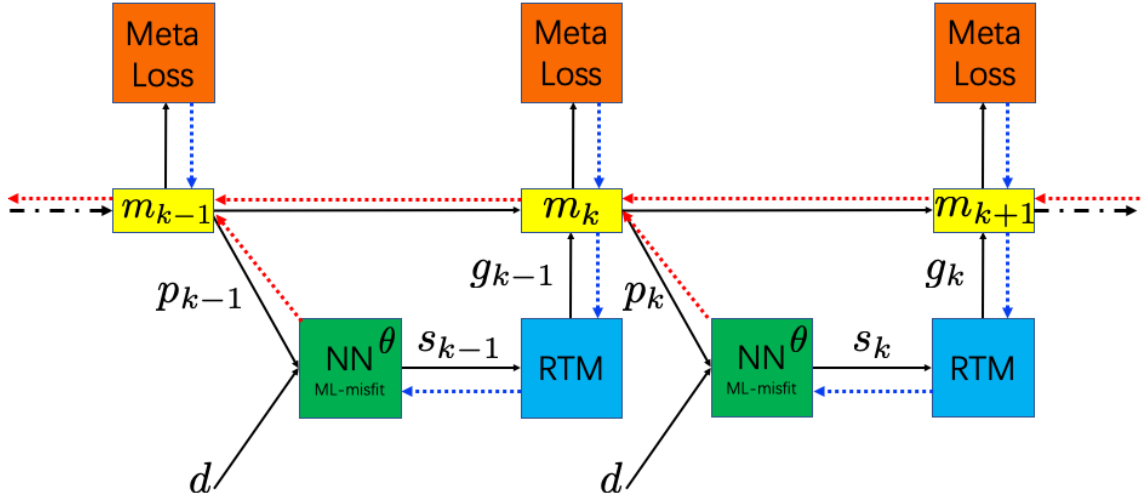
\plot{fwi_flow}{width=0.99\textwidth}{A schematic plot describing the flow for the FWI using the ML-misfit function. The green block is the neural network with parameter $\theta$ representing the ML-misfit function. The blue block represents the Reverse Time Migration (RTM) operator. The red block refers to computation of the meta-loss defined by equation \ref{eqn_meta_loss}. The yellow block denotes the velocity model $m_k$. $p_k$ is the predicted data modeled by the velocity model and $d$ is the measured datasets. $s_k$ is the adjoint source computed from the ML-misfit. $g_k$ is the gradient output from the RTM block given an adjoint source $s_k$. The blue and red curves denote the gradient flow for updating parameters $\theta$ of the ML-misfit. The blue curves are related to the first-order dependency between the meta-loss and the parameters $\theta$, while red curves correspond to higher-order dependencies. }
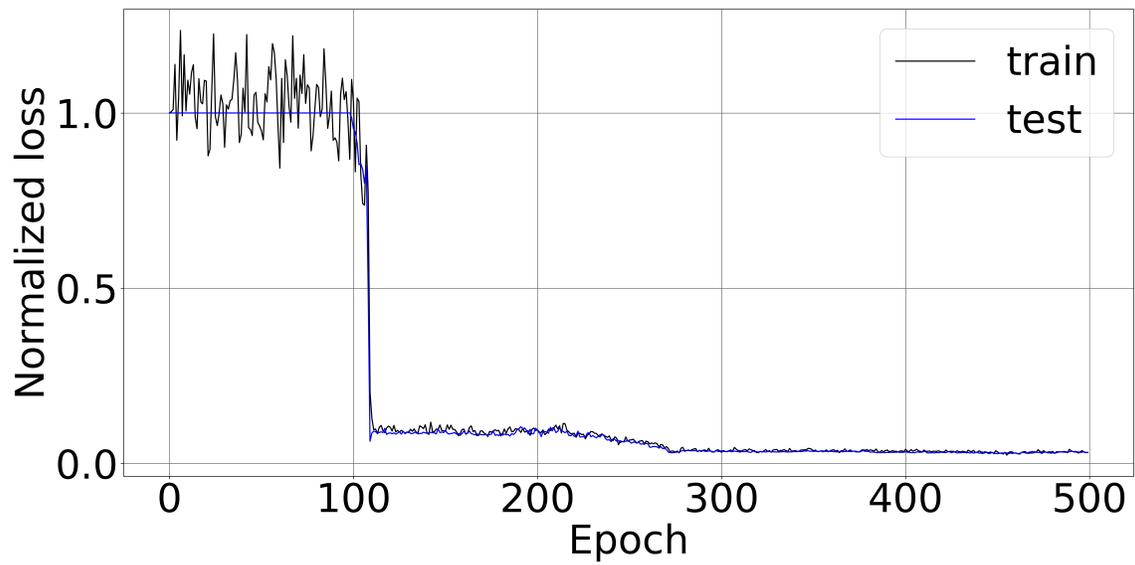
\plot{misfit}{width=0.99\textwidth}{The loss over epochs for training the ML-misfit of the travel time-shifted signals example. The loss is the $l_2$ norm of the difference between the inverted travel time and the true travel time. }
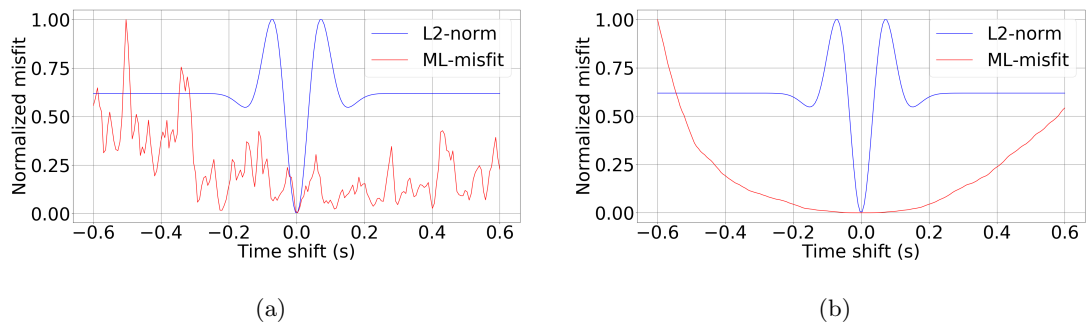
\multiplot{2}{convexity_without_train,convexity_train_500}{width=0.45\textwidth}{The misfit curves for the $l_2$ norm and the ML-misfit with respect to time shifts: ML-misfit a) without training; b) training over 500 epochs.}

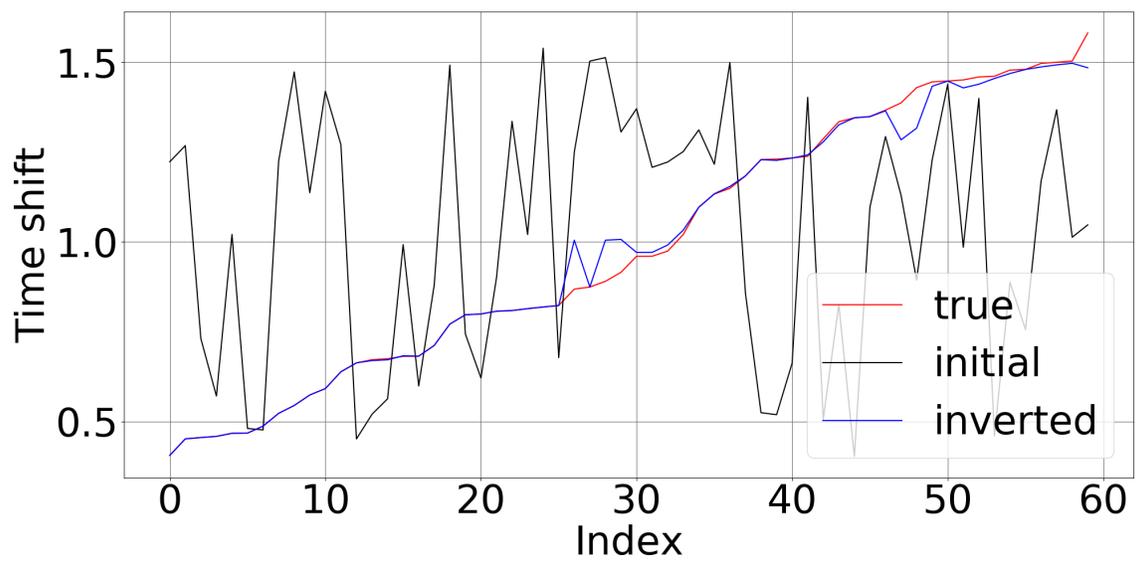
\plot{test_check}{width=0.99\textwidth}{The inverted result for the testing dataset using the trained ML-misfit function.}
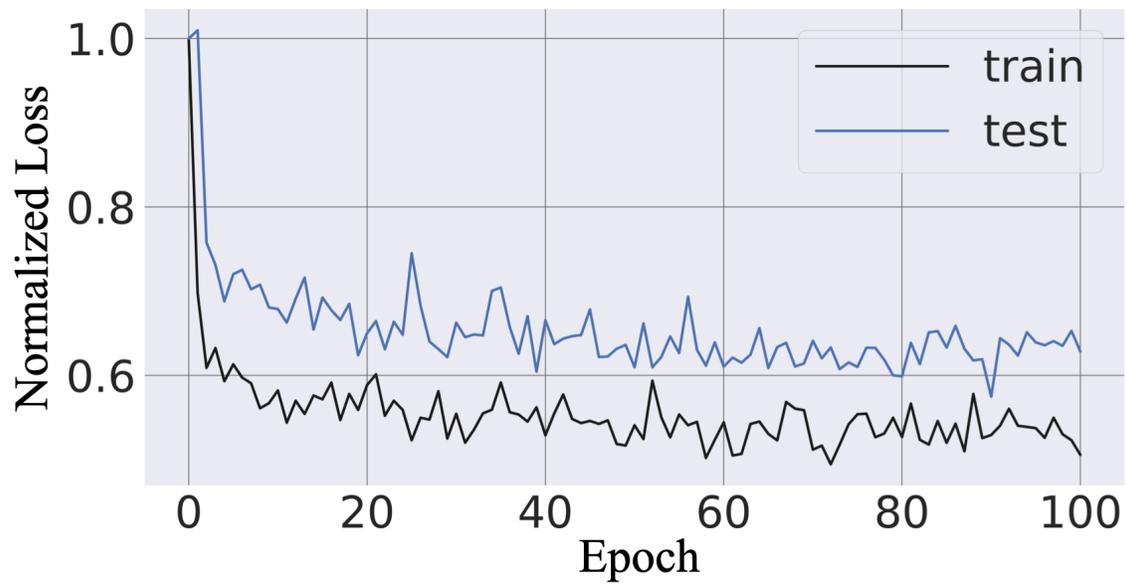
\plot{marm_epoch}{width=0.99\textwidth}{The loss over epochs for the training (black curve) and testing (blue curve) tasks. }
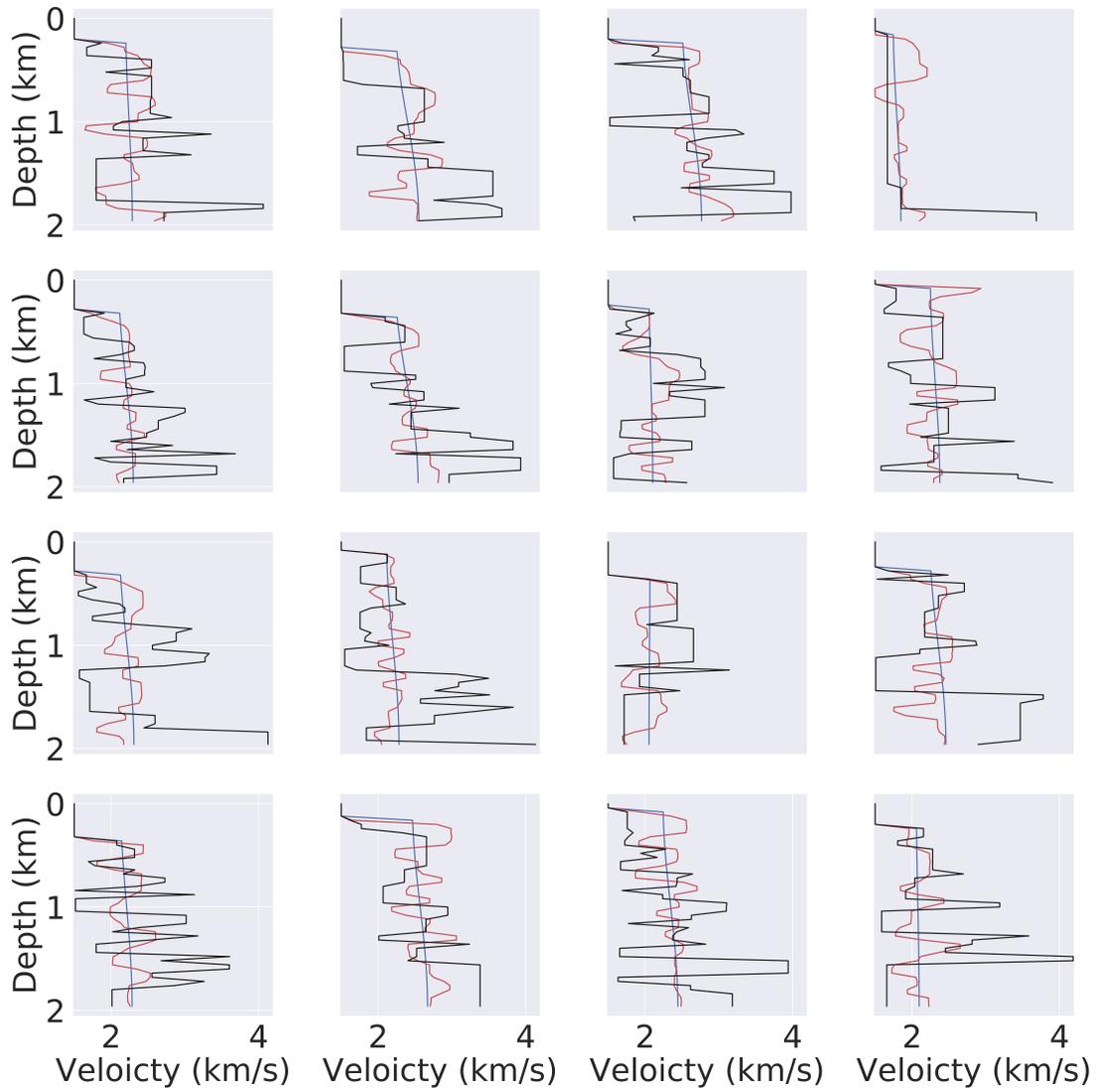
\plot{NEW_check0False}{width=0.99\textwidth}{The velocity profiles for the inversion of the testing tasks using the ML-misfit (without training). Blue curves correspond to initial models, black curves correspond to the true models while red curves correspond to inverted models. }
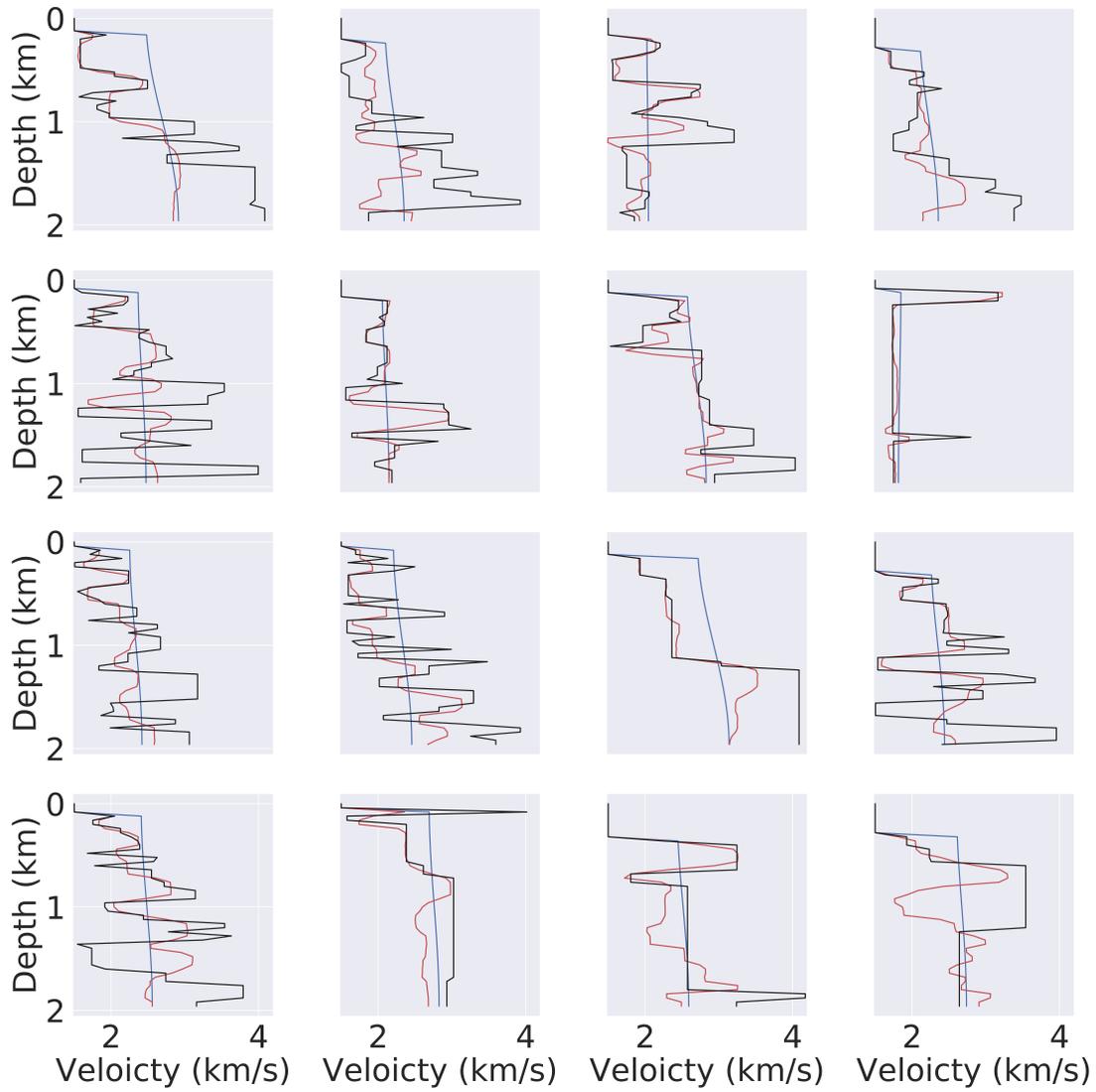
\plot{NEW_check110False}{width=0.99\textwidth}{The velocity profiles for the inversion of the testing tasks using the ML-misfit (after training). Blue curves correspond to initial models, black curves correspond to the true models while red curves correspond to inverted models. }
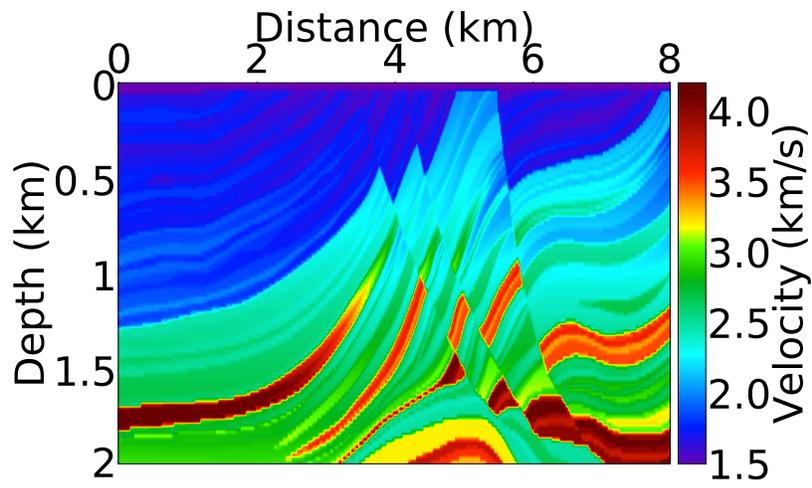
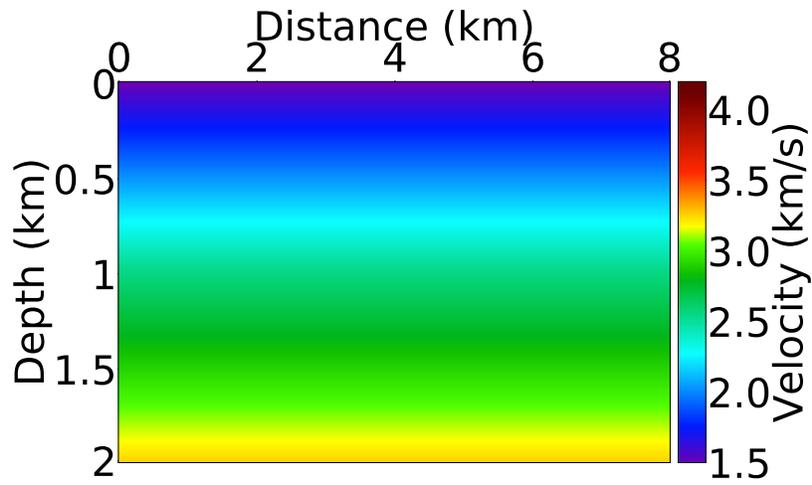
\multiplot{2}{marm_vtrue2,marm_vini2}{width=0.70\textwidth}{a) The true velocity model corresponding to the modified Marmousi and b) the initial velocity.}
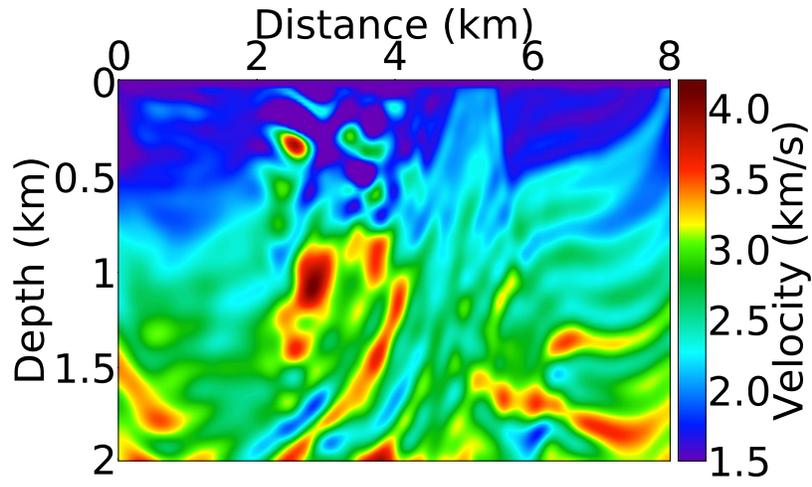
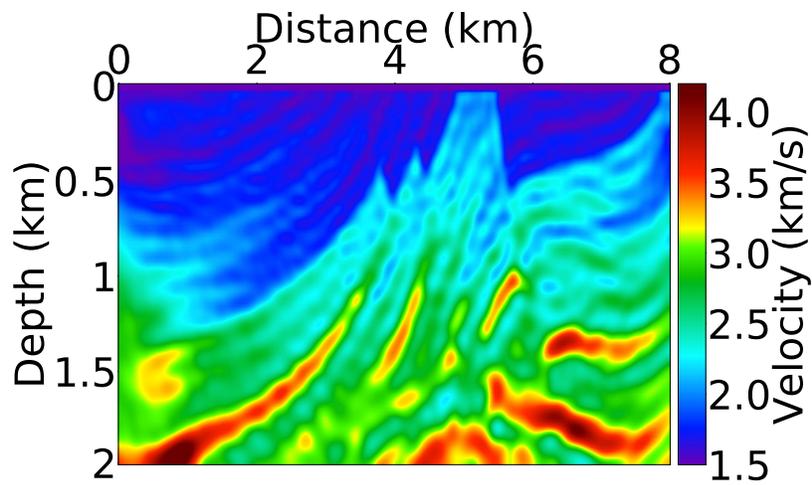
\multiplot{2}{marm_L2_misfit,marm_ML_misfit}{width=0.70\textwidth}{The inverted model by a) the $l_2$ norm misfit function; b) the ML-misfit. Note that the initial model for starting inversion is shown in Figure \ref{fig:marm_vini2} and the true model is shown in Figure \ref{fig:marm_vtrue2}.}
%\multiplot{9}{convexityaa_2_epoch0,convexityaa_2_epoch100,convexityaa_2_epoch200,convexityaa_2_epoch300,convexityaa_2_epoch400,convexityaa_2_epoch500,convexityaa_2_epoch600,convexityaa_2_epoch700,convexityaa_2_epoch800}{width=0.30\textwidth}{Convexity analysis for the ML-misfit and L2 norm misfit. The ML-misfit function after a) 1; b) 100 ; c) 200; d) 300; e) 400; f) 500; g) 600; h) 700; i) 800 epoch training. Note, in training the ML-misfit, the Hinge loss function is included.}
%\multiplot{9}{convexityaa_0_epoch0,convexityaa_0_epoch100,convexityaa_0_epoch200,convexityaa_0_epoch300,convexityaa_0_epoch400,convexityaa_0_epoch500,convexityaa_0_epoch600,convexityaa_0_epoch700,convexityaa_0_epoch800}{width=0.30\textwidth}{Convexity analysis for the ML-misfit and L2 norm misfit. The ML-misfit function after a) 1; b) 100 ; c) 200; d) 300; e) 400; f) 500; g) 600; h) 700; i) 800 epoch training. Note, in training the ML-misfit, the Hinge loss function is excluded.}

\begin{algorithm}
\caption{Training of the ML-misfit}
\begin{algorithmic}[1] 
\State Initialize the ML-misfit neural network $\Phi(p,d;\theta)$
\State Initialize the learning rate $\gamma$ for FWI 
\State Initialize the learning rate $\eta$ for meta-learning 
\State Initialize the unroll parameter K 
\For{${episode=1,\text{M}}$}
\State Random select true model $v_\text{true}$, initial model $v_{ini}$
\State Compute the measured data $d=\text{modeling}(v_\text{true})$
\State Initialize current model $v=v_{ini}$ 
\State Initialize the accumulated meta-loss $L = 0 $ 
\For{${t=1,T}$}
\State Compute the predicted data $p=\text{modeling}(v)$
\State Compute the adjoint source for ML-misfit $s = \frac{\partial \Phi(p,d;\theta)}{\partial p}$
\State Compute the gradient $g_v=\text{RTM}(s)$
\State Update the model $v = v - \gamma *g_v$
\If{$t$ mod K $==0$}
\State Compute the derivative of $L$ with respect to $\theta$ with auto-differentiation: $g_{\theta}$
\State Update NN parameter $\theta= \theta -\eta g_{\theta}$
\State Reset the accumulated meta-loss $L = 0$
\State Detach the current model $v$ from the computation graph 
\Else
\State Compute the meta-loss $L_t$ for current step by equation \ref{eqn_meta_loss}
\State Accumulate the meta-loss $L = L + L_t$
\EndIf
\EndFor
\EndFor
\end{algorithmic}
\end{algorithm}

\append{The Optimal transport of the matching filter misfit for FWI}
The conventional $l_2$ norm misfit defined in equation \ref{misfit_l2} seeks a local point-wise comparison between the predicted data $p(t)$ and the measured data $d(t)$. Such local comparison makes it prone to cycle-skipping. \cite{MF_OTMF} introduced the optimal transport of the matching filter (OTMF) misfit for FWI. In OTMF, a matching filter is computed first by deconvolving the predicted data from the measured ones:
\begin{equation}
d(t)*w(t)=p(t),
\label{eqn_deconv}
\end{equation}
where $*$ denotes the convolution operation. The resulting matching filter $w(t)$ is further preconditioned to be a density function of a probability distribution $\tilde{w}(t)$:
\begin{equation}
\tilde{w}(t) = \frac{w^2(t)}{||w(t)||_2^2}.
\end{equation}
When the subsurface velocity model is correct, the measured data $d(t)$ and the predicted ones $p(t)$ are identical leading to a Dirac delta function for the matching filter distribution $\tilde{w}(t)$. Thus, the OTMF misfit function is defined as the Wasserstein $W_2$ distance between the resulting matching filter distribution and a target Dirac delta function:
\begin{equation}
J_{\text{OTMF}} = W_2^2(\tilde{w}(t),\delta(t)),
\label{eqn_OTMF}
\end{equation}
where $W_2$ denotes the Wasserstein distance \citep{opt2}. The resulting OTMF misfit in Equation 3 can overcome cycle-skipping effectively as demonstrated by \cite{SUN_OTMF_EAGE}. For an in-detail discussion of the properties of the OTMF misift, such as the its convexity, adjoint source computation and comparison with other misfits, we refer you to \cite{MF_OTMF}. Specially, if we consider the Wasserstein $W_2$ distance between two Gaussian distributions, we obtain \citep{OLKIN1982257}:
\begin{equation}
W_2^2\left[(\mu,\sigma),(\mu_0,\sigma_0)\right] = ||m-m_0||_2^2=(\mu-\mu_0)^2 + (\sigma-\sigma_0)^2 
\label{eqn_gauss}
\end{equation}
where $\mu,\mu_0$ and $\sigma,\sigma_0$ are the means and variances of the Gaussian distributions, respectively, $m=[\mu,\sigma]$ and $m_0=[\mu_0,\sigma_0]$ are vectors constituted by the mean and variance. Thus, we can interpret the OTMF misfit function of equation \ref{eqn_OTMF} as the L2 norm between the statistics of the matching filter distribution $\tilde{w}(t)$ and the delta function $\delta (t)$:
\begin{equation}
J_{OTMF} =  ||m_{\tilde{w}(t)} -m_{\delta(t)} ||_2^2
\label{OTMF_stat}
\end{equation}
Here, we emphasize the structure of the OTMF misfit of equation \ref{eqn_OTMF} as it inspires us to design the form of the ML-misfit in equation \ref{ML_misfit_form}:
\begin{itemize}
	\item First, there is a deconvolution step of equation \ref{eqn_deconv} involved in the OTMF misfit. The deconvolution between the predicted $p$ and measured data $d$ provides a global comparison and helps to overcome cycle-skipping. Thus, the introduced neural network should take the predicted data $p$ and the observed data $d$ together as input to mimic this process. As a result, we have $\phi(p,d;\theta)$ in the first term of equation \ref{ML_misfit_form}.
	\item We also include the term $\phi(d,d;\theta)$ to mimic the role of the target delta function in equation A-3.
	\item As the Wasserstein distance can be explained as the $l_2$ norm between the statistics of the distributions as in equation \ref{OTMF_stat},  we include the $l_2$ norm for comparison of the term containing $\phi(p,d;\theta)$ and $\phi(d,d;\theta)$ in the first term of equation \ref{eqn_OTMF} and also because the term containing $\phi(p,d;\theta)$ and $\phi(p,d;\theta)$ is expected to be the a vector of the statistical measure of resulting distributions, the length of those vectors would be at least 2 (like the mean and variance for a Gaussian distribution).
\end{itemize}

\bibliography{../../mypaper}
\end{document}